\documentclass[reprint,
 amsmath,amssymb,
 aps,
]{revtex4-2}

\usepackage{graphicx}
\usepackage{dcolumn}
\usepackage{bm}
\usepackage{xcolor}

\begin{document}

\title{Non-differentiable angular dispersion as an optical resource}

\author{Layton A. Hall$^{1}$}
\author{Ayman F. Abouraddy$^{1,*}$}
\affiliation{$^{1}$CREOL, The College of Optics \& Photonics, University of Central~Florida, Orlando, FL 32816, USA}
\affiliation{$^*$Corresponding author: raddy@creol.ucf.edu}

\begin{abstract}
Introducing angular dispersion into a pulsed field associates each frequency with a particular angle with respect to the propagation axis. A perennial yet implicit assumption is that the propagation angle is differentiable with respect to the frequency. Recent work has shown that the existence of a frequency at which the derivative of the propagation angle does not exist -- which we refer to as non-differentiable angular dispersion -- allows for the optical field to exhibit unique and useful characteristics that are unattainable by endowing optical fields with conventional angular dispersion. Because these novel features are retained in principle even when the specific non-differentiable frequency is not part of the selected spectrum, the question arises as to the impact of the proximity of the spectrum to this frequency. We show here that operating in the vicinity of the non-differentiable frequency is imperative to reduce the deleterious impact of (1) errors in implementing the angular-dispersion profile, and (2) the spectral uncertainty intrinsic to finite-energy wave packets in any realistic system. Non-differential angular dispersion can then be viewed as a resource -- quantified by a Schmidt number -- that is maximized in the vicinity of the non-differentiable frequency. These results will be useful in designing novel phase-matching of nonlinear interactions in dispersive media. 
\end{abstract}


\maketitle

\section{Introduction}

\begin{figure}[t!]
\centering
\includegraphics[width=8.6cm]{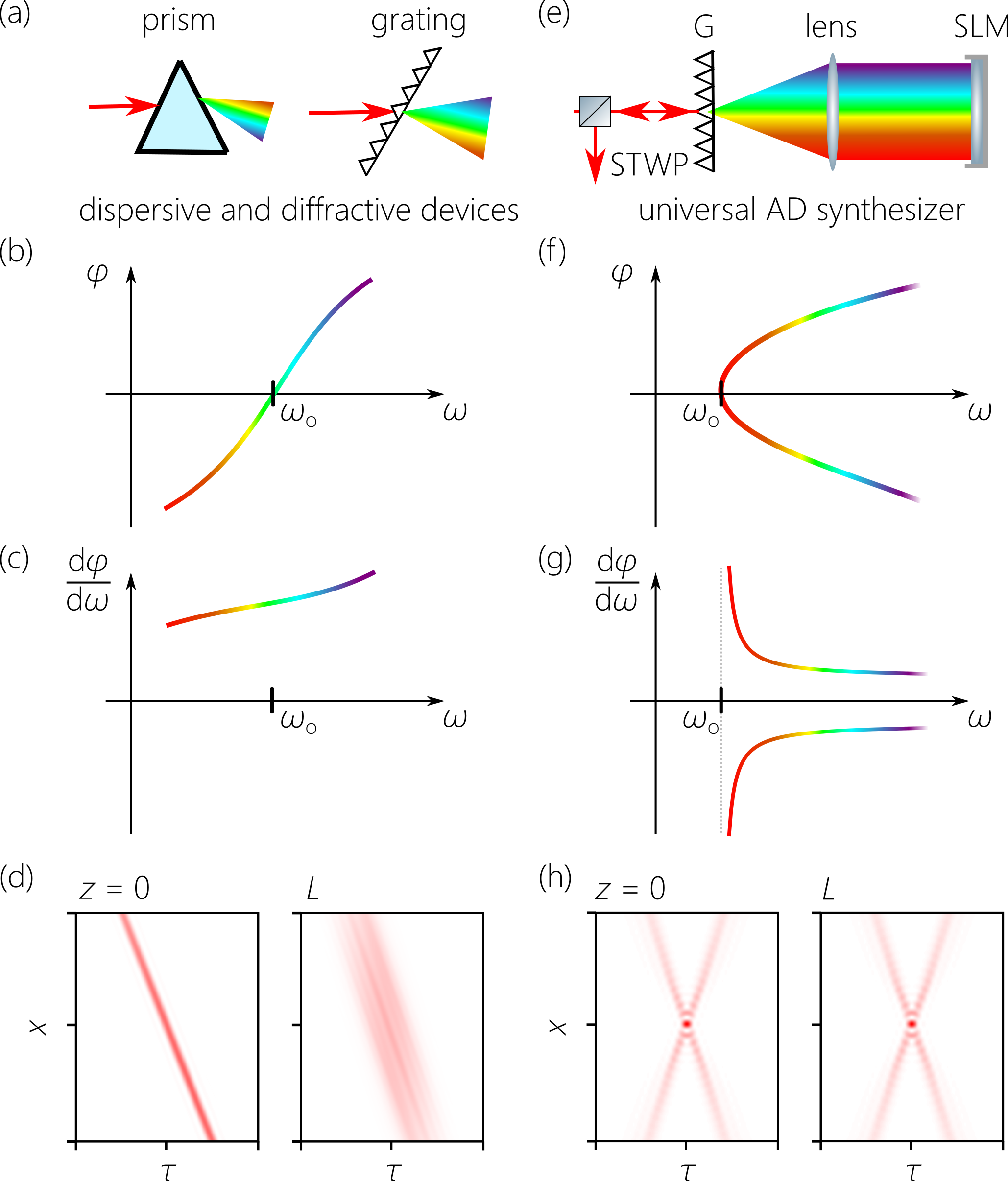}
\caption{(a) Conventional AD is introduced via dispersive or diffractive devices; (b) the propagation angle $\varphi(\omega)$; (c) $\tfrac{d\varphi}{d\omega}$ is finite everywhere; and (d) the resulting TPF experiences GVD in free space. Here the temporal width of the spatio-temporal intensity profile $I(x,z;\tau)$ of the TPF increases at $z\!=\!l$ with respect to that at $z\!=\!0$. (e) Non-differentiable AD is introduced via a universal AD synthesizer \cite{Hall22OE}; (f) $\varphi(\omega)$ is continuous; (g) but $\tfrac{d\varphi}{d\omega}$ is not defined at $\omega_{\mathrm{o}}$; and (h) the resulting STWP is propagation invariant.}
\label{Fig:Concept}
\end{figure}

Angular dispersion (AD) is a venerable subject in optics, extending back to Newton's experiments spectrally resolving white light \cite{Sabra81Book}. It is introduced into optical fields by diffractive or dispersive devices such as gratings or prisms [Fig.~\ref{Fig:Concept}(a)], whereby each wavelength travels at a different angle with respect to the propagation axis \cite{Torres10AOP}; see Fig.~\ref{Fig:Concept}(b,c). In the case of a coherent pulse, AD tilts the pulse front (the plane of constant amplitude) with respect to the phase front (the plane of constant phase), and such a pulse is then called a `tilted pulse front' (TPF) \cite{Fulop10Review}; see Fig.~\ref{Fig:Concept}(d). Changing the pulse-front tilt tunes the group velocity \cite{Hebling02OE} and the group-velocity dispersion (GVD) \cite{Szatmari96OL} experienced by the TPF in free space. This makes TPFs useful in a broad range of optical applications, including traveling-wave optical amplification \cite{Bor83APB,Klebniczki88APB,Hebling89OL,Hebling91JOSAB}, dispersion compensation \cite{Fork84OL,Gordon84OL,Szatmari96OL}, pulse compression \cite{Bor85OC,Lemoff93OL,Kane97JOSAB,Kane97JOSAB2}, broadening the bandwidth of phase-matching in nonlinear optics \cite{Martinez89IEEE,Szabo90APB,Szabo94APB,Richman98OL,Richman99AO}, and in generating THz pulses \cite{Hebling02OE,Nugraha19OL,Wang20LPR}.
 
Until recently, the established theory of AD and TPFs appeared complete \cite{Hebling96OQE,Fulop10Review,Torres10AOP}. However, investigations of a new class of pulsed beams dubbed `space-time wave packets' (STWPs) \cite{Kondakci16OE,Parker16OE,Yessenov22AOP} have overturned this widely accepted notion, which has thus opened up a new perspective on AD and its potential applications. STWPs exhibit several unique and desirable attributes such as propagation invariance in linear media \cite{Malaguti08OL,Malaguti09PRA,FigueroaBook14,Kondakci17NP,Porras17OL,Kondakci18PRL,Bhaduri19Optica,Wong20AS,Guo21Light,Bejot22ACSP}, their group velocity is tunable over an unprecedented scale \cite{Salo01JOA,Wong17ACSP2,Efremidis17OL,Kondakci19NC,Yessenov19OE,Li20CP}, they exhibit anomalous refraction effects \cite{Bhaduri20NP,Motz21OL}, self-healing \cite{Kondakci18OL}, Talbot effects in space-time \cite{Hall21APLP}, and axial self-acceleration \cite{Yessenov20PRL2,Li20SR,Li21CP,Hall22OL}. Moreover, STWPs interact with photonic devices such as waveguides \cite{Shiri20NC,Kibler21PRL,Guo21PRR,Bejot21ACSP,Shiri22Optica} and planar cavities \cite{Shiri20OL} in novel ways. Although AD undergirds STWPs [Fig.~\ref{Fig:Concept}(e-h)] just as in the case of TPFs \cite{Wong17ACSP2,Kondakci19ACSP}, the characteristic features of STWPs nevertheless directly contradict those exhibited by TPFs. For example, whereas TPFs are \textit{always} dispersive in free space [Fig.~\ref{Fig:Concept}(d)], STWPs can be GVD-free [Fig.~\ref{Fig:Concept}(h)]; and whereas the \textit{on-axis} group velocity of a TPF in free space is always $c$ (the speed of light in vacuum), that of a STWP can be tuned arbitrarily \cite{Hall22OE}. Furthermore, a well-known theorem shows that AD can produce only anomalous effective GVD in free space \cite{Martinez84JOSAA}, whereas both normal and anomalous GVD can be exhibited by an STWP \cite{Yessenov21ACSP,Hall21OLnormalGVD}.

Why are the defining characteristics of STWPs in such direct contradistinction to those of TPFs, even though AD undergirds both classes of fields? Our recent investigations \cite{Hall21OL,Hall21OLnormalGVD,Hall22OE} have identified the source of this variance: whereas the AD introduced by conventional optical components (prisms, gratings, and even metasurfaces \cite{Arbabi17Optica,McClung20Light}) into TPFs is `differentiable' [Fig.~\ref{Fig:Concept}(b,c)], STWPs are endowed instead with `non-differentiable' AD [Fig.~\ref{Fig:Concept}(f,g)]. By differentiable AD we mean that the propagation angle $\varphi(\omega)$ is a differentiable function with respect to the temporal frequency $\omega$; i.e., $\tfrac{d\varphi}{d\omega}$ and higher-order derivatives are finite and $\varphi(\omega)$ can therefore be expanded perturbatively at any frequency into a Taylor series. It may initially appear that the \textit{non-differentiability} of the AD undergirding STWPs is thus an exotic feature that is difficult to produce in practice. This is not the case [Fig.~\ref{Fig:Concept}(e)]. By non-differentiable AD we mean that the propagation angle $\varphi(\omega)$ is continuous but does \textit{not} possess a derivative at some frequency $\omega_{\mathrm{o}}$, which we refer to as the non-differentiable frequency [Fig.~\ref{Fig:Concept}(f)]. At $\omega_{\mathrm{o}}$, $\tfrac{d\varphi}{d\omega}$ and higher-order derivatives are \textit{not} defined [Fig.~\ref{Fig:Concept}(g)], so that $\varphi(\omega)$ can\textit{not} be expanded perturbatively into a Taylor series at $\omega_{\mathrm{o}}$.

Crucially, even if the STWP spectrum is truncated so as not to include the non-differentiable frequency $\omega_{\mathrm{o}}$ itself, and $\varphi(\omega)$ is thus differentiable everywhere in the bandwidth of interest, the unique consequences of non-differentiable AD are nevertheless -- in principle -- maintained. A crucial question therefore arises with regards to the effect of the proximity of the STWP spectrum to the non-differentiable frequency $\omega_{\mathrm{o}}$. Is it useful for the STWP spectrum to be selected in the vicinity of $\omega_{\mathrm{o}}$?

We show here that it is extremely useful to operate in the vicinity of the non-differentiable frequency. We demonstrate that two major advantages motivate this conclusion. First, when the STWP spectrum is in the vicinity of $\omega_{\mathrm{o}}$, the tolerance is much higher with regards to imprecision or errors in implementing the desired AD profile $\varphi(\omega)$. In contrast, operating far from $\omega_{\mathrm{o}}$ imposes strict tolerances on the realized AD profile that must be satisfied to avoid having the field attributes diverge significantly from those targeted. Second, we show that operating in the vicinity of the non-differentiable frequency $\omega_{\mathrm{o}}$ reduces the deleterious impact of `spectral uncertainty', which refers to the inevitable \textit{finite} spectrum associated with each propagation angle $\varphi$ in any realistic system \cite{Yessenov19OE}. Operating far from $\omega_{\mathrm{o}}$ in presence of spectral uncertainty reduces the accessible control over the STWP group velocity and its dispersion characteristics.

We therefore view the non-differentiability of AD as an optical resource that can be exploited optimally by operating in the vicinity of $\omega_{\mathrm{o}}$. We quantify this resource in terms of the Schmidt number for the spatio-temporal spectrum, which increases as the spectrum approaches the non-differentiable frequency even in presence of spectral uncertainty, indicating the large number of underlying spatio-temporal modes. As the spectrum moves away from $\omega_{\mathrm{o}}$, the Schmidt number drops, and therefore only a few modes contribute to the spectrum, which reduces the extent of the control that can be exercised over the STWP properties.

\section{Consequences of non-differentiable angular dispersion}

\begin{figure}[t!]
\centering
\includegraphics[width=8.6cm]{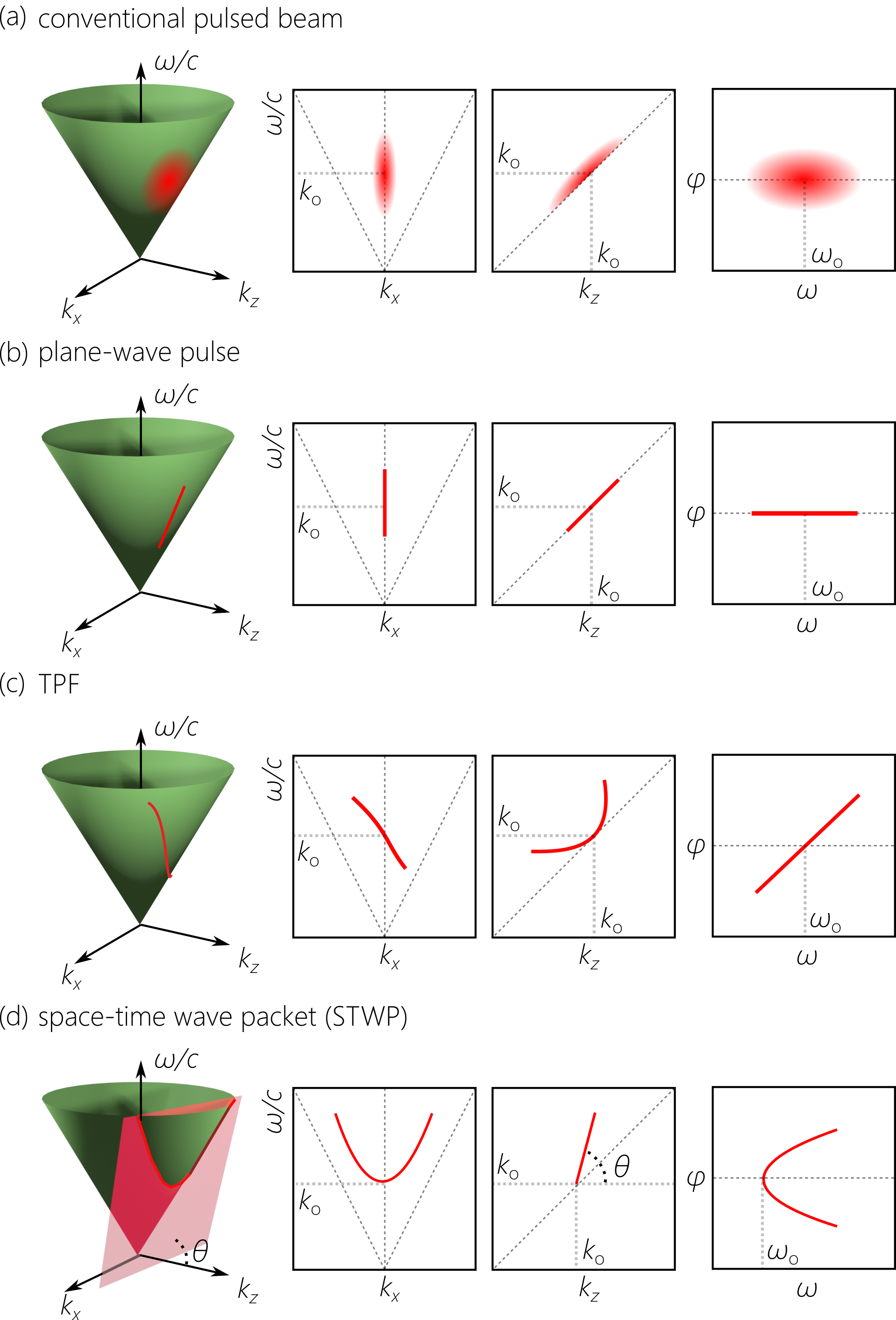}
\caption{(a) The spectral support on the light-cone surface for a conventional pulsed beam, along with the spectral projections onto the $(k_{x},\tfrac{\omega}{c})$ and $(k_{z},\tfrac{\omega}{c})$ planes, in addition to the propagation angle $\varphi(\omega)$. (b) Same as (a) for a plane-wave pulse; (c) a TPF after introducing differentiable AD; and (d) an STWP after introducing non-differentiable AD.}
\label{Fig:LightCones}
\end{figure}

We consider scalar optical fields with one transverse spatial dimension $x$ and hold the field uniform along $y$, as is usual in studies of AD. A conventional pulsed beam $E(x,z;t)\!=\!e^{i(k_{\mathrm{o}}z-\omega_{\mathrm{o}}t)}\psi(x,z;t)$ has a slowly varying envelope that can be expressed in terms of an angular spectrum as follows:
\begin{equation}\label{Eq:GeneralEnvelope}
\psi(x,z;t)=\iint\!dk_{x}d\Omega\,\widetilde{\psi}(k_{x},\Omega)e^{ik_{x}x}e^{i(k_{z}-k_{\mathrm{o}})z}e^{-i\Omega t},
\end{equation}
where $\omega_{\mathrm{o}}$ is a carrier frequency, $k_{\mathrm{o}}\!=\!\omega_{\mathrm{o}}/c$ the associated wave number, $\Omega\!=\!\omega-\omega_{\mathrm{o}}$, $k_{x}$ the transverse wave number (or spatial frequency), $k_{z}\!=\!\sqrt{(\tfrac{\omega}{c})^{2}-k_{x}^{2}}$ is the axial wave number, and the spatio-temporal spectrum $\widetilde{\psi}(k_{x},\Omega)$ is the Fourier transform of $\psi(x,0;t)$. The spectral support for such a generic wave packet on the surface of the light-cone $k_{x}^{2}+k_{z}^{2}\!=\!(\tfrac{\omega}{c})^{2}$ \cite{Donnelly93ProcRSLA} is in general a 2D domain [Fig.~\ref{Fig:LightCones}(a)]. Because of the spectral extension along $k_{x}$ and $\omega$, such a pulsed beam does not have a well-defined propagation-angle profile $\varphi(\omega)$. In the simple case of a pulsed \textit{plane wave} traveling along $z$, the envelope is independent of $x$ and propagates invariantly at a group velocity $\widetilde{v}\!=\!c$:
\begin{equation}\label{Eq:PlaneWavePulse}
\psi(x,z;t)=\psi(z;t)=\int\!d\Omega\,\widetilde{\psi}(\Omega)e^{-i\Omega(t-z/c)}=\psi(0;t-z/c),
\end{equation}
where $\widetilde{\psi}(\Omega)$ is the Fourier transform of $\psi(0;t)$ [Fig.~\ref{Fig:LightCones}(b)].

After this plane-wave pulse traverses a dispersive or diffractive device, each frequency $\omega$ travels at a different angle $\varphi(\omega)$ with the $z$-axis, so that:
\begin{equation}\label{Eq:PulseWithAD}
\psi(x,z;t)=\int\!d\Omega\,\widetilde{\psi}(\Omega)e^{i\{k_{x}(\omega)x+(k_{z}(\omega)-k_{\mathrm{o}})z-\Omega t\}},
\end{equation}
where $k_{x}(\omega)\!=\!\tfrac{\omega}{c}\sin{\{\varphi(\omega)\}}$ and $k_{z}(\omega)\!=\!\tfrac{\omega}{c}\cos{\{\varphi(\omega)\}}$, which corresponds in general to a TPF. The spectral support for a TPF is thus a 1D trajectory on the light-cone surface parameterized by $\varphi(\omega)$. In general, any pulsed field incorporating AD will be represented by such a 1D spectral support [Fig.~\ref{Fig:LightCones}(c)]. The group velocity $\widetilde{v}$ of a TPF along the $z$-axis is given by \cite{Porras03PRE2}:
\begin{equation}
\widetilde{v}=\frac{c}{\cos{\varphi_{\mathrm{o}}}-\omega_{\mathrm{o}}\varphi_{\mathrm{o}}^{(1)}\sin{\varphi_{\mathrm{{o}}}}},
\end{equation}
where $\varphi_{\mathrm{o}}\!=\!\varphi(\omega_{\mathrm{o}})$ and $\varphi_{\mathrm{o}}^{(1)}\!=\!\tfrac{d\varphi}{d\omega}\big|_{\omega_{\mathrm{o}}}$. Of course, if the TPF propagates along the $z$-axis, $\varphi_{\mathrm{o}}\!=\!0$ and consequently we have $\widetilde{v}\!=\!c$. Only when $\varphi_{\mathrm{o}}\!\neq\!0$ can the group velocity of a TPF along the $z$-axis deviate from $c$.

STWPs are also undergirded by AD, so their spectral support on the light-cone is also a 1D trajectory. However, the spectral support for a propagation-invariant STWP takes the form of a conic section at the intersection of the light-cone with the plane $\Omega\!=\!(k_{z}-k_{\mathrm{o}})c\tan{\theta}$ [Fig.~\ref{Fig:LightCones}(d)]. The plane is parallel to the $k_{x}$-axis and makes an angle $\theta$ (the spectral tilt angle) with the $k_{z}$-axis \cite{Kondakci17NP,Yessenov22AOP}. The envelope now takes the form:
\begin{equation}\label{Eq:SWTPEnvelope}
\psi(x,z;t)=\int\!d\Omega\,\widetilde{\psi}(\Omega)e^{ik_{x}(\omega)x}e^{-i\Omega(t-z/\widetilde{v})}=\psi(x,0;t-z/\widetilde{v}),
\end{equation}
which indicates that the STWP is propagation invariant and travels rigidly in free space at a group velocity $\widetilde{v}\!=\!c\tan{\theta}$ \cite{Kondakci19NC}. The group velocity can thus take on -- in principle -- an arbitrary value, whether subluminal $\widetilde{v}\!<\!c$ ($0^{\circ}\!<\!\theta\!<\!45^{\circ}$), superluminal $\widetilde{v}\!>\!c$ ($45^{\circ}\!<\!\theta\!<\!90^{\circ}$), or negative $\widetilde{v}\!<\!0$ ($90^{\circ}\!<\!\theta\!<\!180^{\circ}$) \cite{Salo01JOA,Efremidis17OL,Yessenov19PRA}. In the paraxial regime, we have $\tfrac{\Omega}{\omega_{\mathrm{o}}}\!\approx\!\tfrac{k_{x}^{2}}{2k_{\mathrm{o}}^{2}(1-\cot{\theta})}$ \cite{Kondakci19NC}, from which we have the small-angle approximation:
\begin{equation}
\varphi(\Omega)\approx\sqrt{2\frac{\Omega}{\omega_{\mathrm{o}}}(1-\cot{\theta})};
\end{equation}
here $\Omega<0$ when $\theta<45^{\circ}$ (subluminal regime), and $\Omega>0$ when $\theta>45^{\circ}$ (superluminal regime). However, because the derivative of $\varphi(\Omega)\!\propto\!\sqrt{\Omega}$ is not defined at $\Omega\!=\!0$, this AD profile is an example of non-differentiable AD, where $\omega\!=\!\omega_{\mathrm{o}}$ is the non-differentiable frequency. The first consequence of this non-differentiable AD profile is that the STWP is propagation invariant as shown in Eq.~\ref{Eq:SWTPEnvelope}. The second consequence is that the on-axis group velocity (along $z$) is now tunable because $\omega\varphi\tfrac{d\varphi}{d\omega}\!\rightarrow\!1-\cot{\theta}$ when $\omega\!\rightarrow\!\omega_{\mathrm{o}}$, so that $\widetilde{v}\!\rightarrow\!c/\widetilde{n}$, where the group index is $\widetilde{n}\!=\!\cot{\theta}$. Therefore, $\widetilde{v}$ can be tuned over an exceptionally broad range (subluminal, superluminal, and negative-$\widetilde{v}$) while remaining on-axis ($\varphi_{\mathrm{o}}\!\rightarrow\!0$) and in the paraxial regime ($\varphi$ is a few degrees). Furthermore, because $\widetilde{v}$ is frequency independent, all higher-order dispersion terms are eliminated.

Therefore, non-differentiability of the AD spectrum at \textit{one} frequency (at least) in an STWP provides the basis for a host of useful propagation characteristics. Of course, one may choose to operate away from this frequency, in which case the AD is differentiable across the bandwidth of interest. Nevertheless, the consequences of non-differentiable AD still ensue in principle. Do any advantages accrue from selecting the STWP spectrum in the vicinity of the non-differentiable frequency $\omega_{\mathrm{o}}$? We proceed to show that there are compelling reasons for operating in the vicinity of the $\omega_{\mathrm{o}}$. Indeed, we garner two crucial advantages: robustness with respect to errors in realizing the target AD profile $\varphi(\omega)$, and retaining control over the group velocity and the dispersion coefficients in the presence of unavoidable spectral uncertainty. Within this conception, non-differentiability of the angular dispersion can thus be viewed as a new resource in optics, and operating in the vicinity of $\omega_{\mathrm{o}}$ helps maximize the utilization of this resource. We proceed to place this notion on a quantitative basis.

\section{Robustness with respect to errors in the angular-dispersion profile}

\begin{figure}[t!]
\centering
\includegraphics[width=8.6cm]{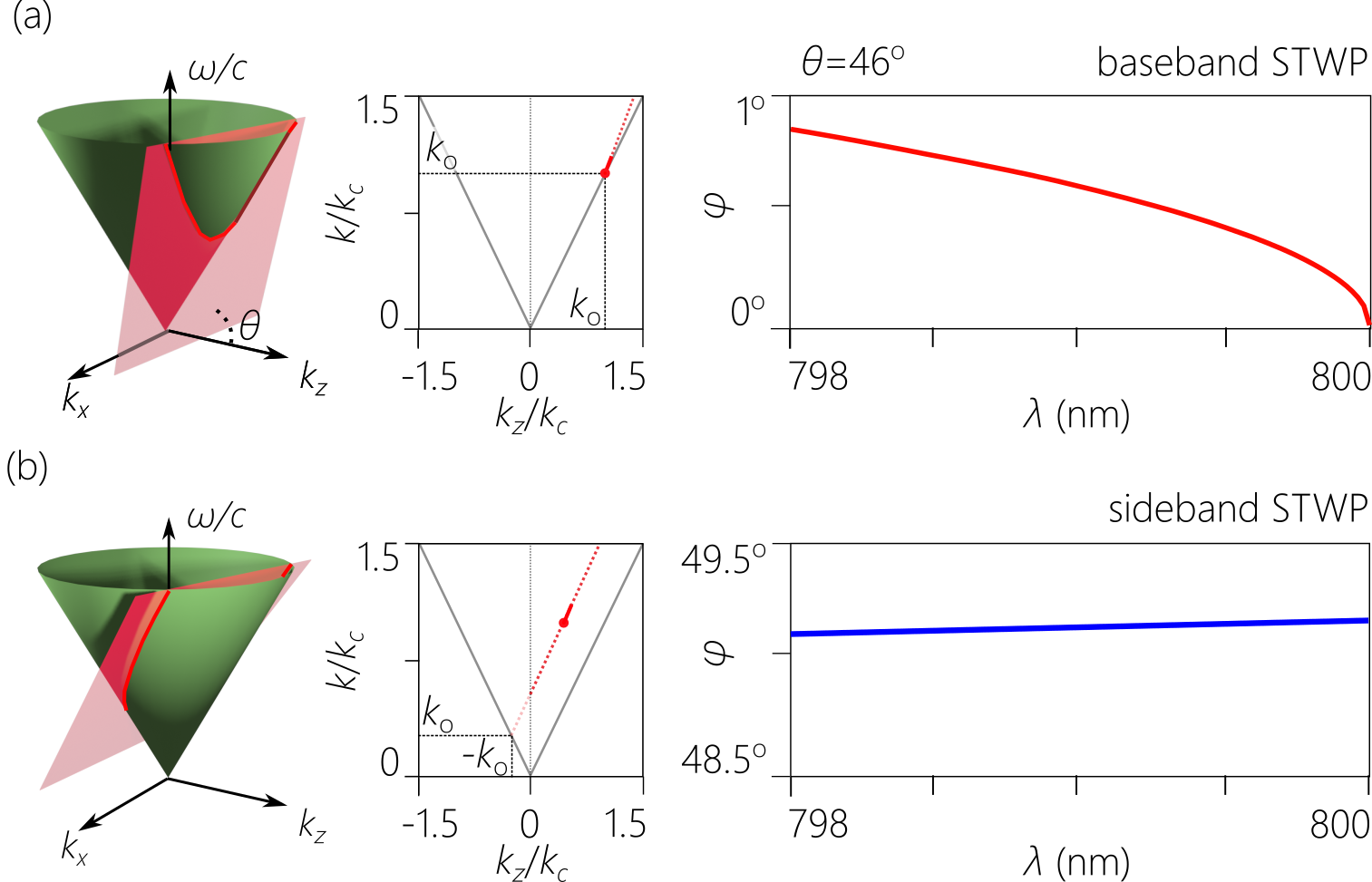}
\caption{Angular-dispersion spectra for (a) a baseband STWP and (b) a sideband STWP. On the left we depict the spectral support for the STWP on the light-cone surface, in the middle we plot the spectral projection onto the $(k_{z},\lambda)$-plane, and on the right we plot $\varphi(\lambda)$. For both STWPs: $\theta\!=\!46^{\circ}$, $\lambda_{\mathrm{c}}\!=\!799$~nm and $\Delta\lambda\!=\!2$~nm. In (a) we have $\lambda_{\mathrm{o}}\!=\!800$~nm, which occurs at $k_{z}\!=\!k_{\mathrm{o}}$ and $\varphi_{\mathrm{o}}\!=\!0$, so that the baseband STWP spectrum includes $\lambda_{\mathrm{o}}$. In (b) we have $\lambda_{\mathrm{o}}\!=\!5$~$\mu$m, which occurs at $k_{z}\!=\!-k_{\mathrm{o}}$ and is thus physically excluded from the spectrum of the sideband STWP.}
\label{Fig:SideBaseAngles}
\end{figure}

\begin{figure}[t!]
\centering
\includegraphics[width=8.6cm]{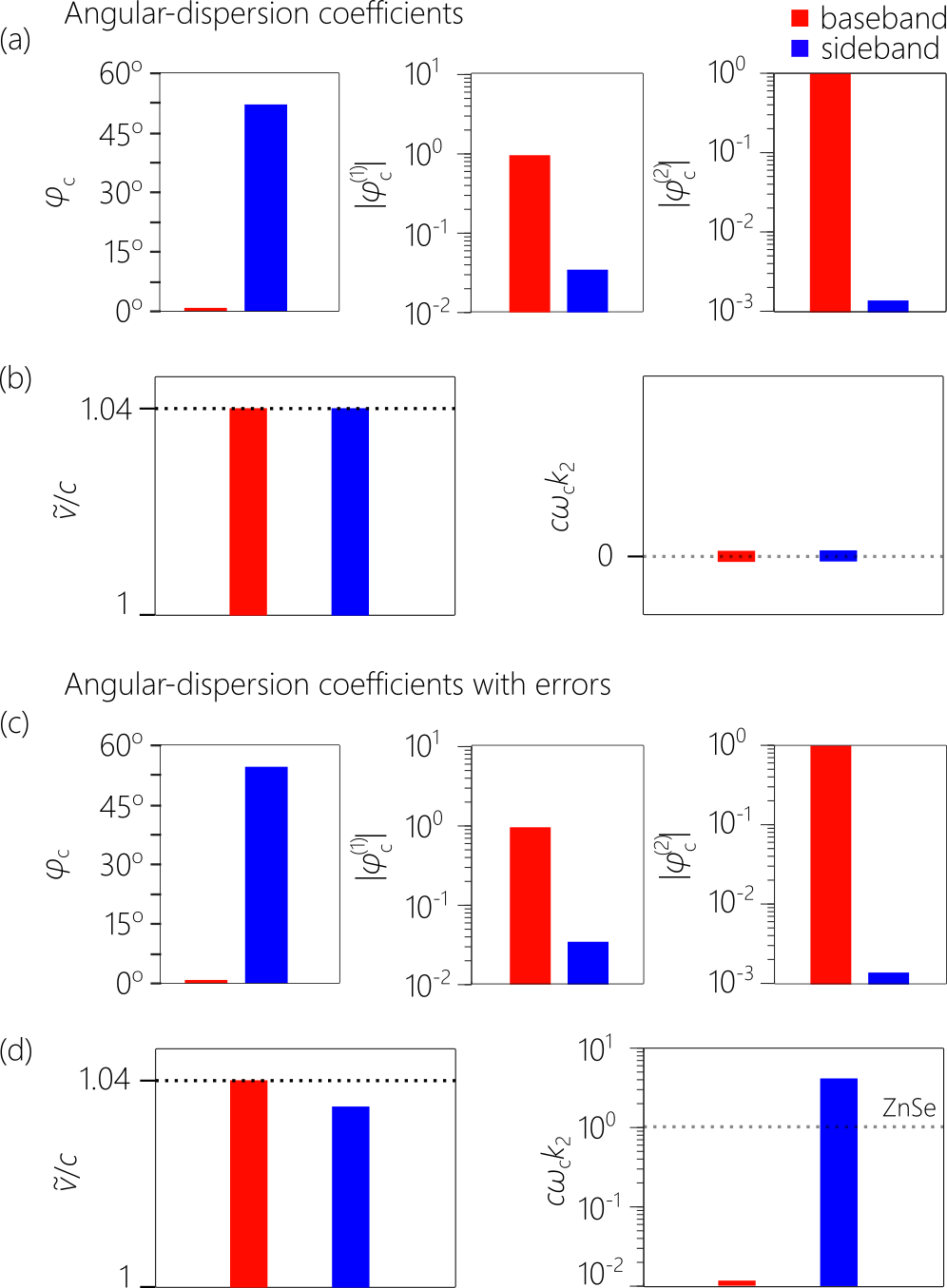}
\caption{Comparing the robustness of sideband and baseband STWPs with respect to implementation errors. (a) The AD coefficients $\varphi_{\mathrm{c}}$, $\varphi_{\mathrm{c}}^{(1)}$, and $\varphi_{\mathrm{c}}^{(2)}$ with out error for the baseband and sideband STWPs from Fig.~\ref{Fig:SideBaseAngles} (the latter two coefficients are normalized with respect to the values of the baseband STWP). (b) Calculated group velocity $\widetilde{v}$ and GVD coefficient $k_{2}$ for the two STWPs. (c) The AD coefficients after introducing a $0.3\%$~error in $\varphi(\omega)$. (d) Calculated $\widetilde{v}$ and $k_{2}$ in presence of the implementation error.}
\label{Fig:Error}
\end{figure}

In any practical setting, there will always be finite errors in implementing a target AD profile $\varphi(\omega)$. We focus here on the impact of such errors on two features of STWPs: their propagation invariance and the tunability of their group velocity $\widetilde{v}$. For this purpose, we compare two classes of STWPs with respect to their performance under errors in the inculcated AD: (1) a `baseband' STWP whose spatial spectrum is centered at $k_{x}\!=\!0$ and includes the non-differentiable frequency $\omega_{\mathrm{o}}$ in its temporal spectrum; and (2) a `sideband' STWP whose spectrum does not include the non-differentiable frequency $\omega_{\mathrm{o}}$ because this frequency is associated with a negative-valued $k_{z}$, which is physically excluded \cite{Yessenov19PRA,Heyman87JOSAA}. The former class of baseband STWPs is the one we have studied exclusively over the past few years \cite{Yessenov22AOP}. On the other hand, sideband STWPs have been investigated for much longer, including focus-wave modes \cite{Brittingham83JAP,Reivelt00JOSAA,Reivelt02PRE}, X-waves \cite{Saari97PRL,Alexeev02PRL,Grunwald03PRA,Reivelt03arxiv,Kuntz09PRA,Bowlan09OL,Bonaretti09OE}, among other examples \cite{FigueroaBook14}. We proceed to show that synthesizing baseband STWPs in the vicinity of the non-differentiable wavelength is dramatically more robust with respect to errors occurring in the realization of the synthesized AD profile in comparison to sideband STWPs, where the non-differentiable wavelength is physically inaccessible. 

Consider a propagation-invariant superluminal STWP with $\theta\!=\!46^{\circ}$ ($\widetilde{v}\!\approx\!1.04c$) and bandwidth $\Delta\lambda\!=\!2$~nm centered at $\lambda_{\mathrm{c}}\!=\!799$~nm. We examine two different implementations: a baseband [Fig.~\ref{Fig:SideBaseAngles}(a)] and a sideband [Fig.~\ref{Fig:SideBaseAngles}(b)] STWP. For the baseband STWP, the spectral support is the intersection of the light-cone with the plane $\Omega\!=\!(k_{z}-k_{\mathrm{o}})\widetilde{v}$, with a non-differentiable wavelength of $\lambda_{\mathrm{o}}\!=\!800$~nm; i.e., $\lambda_{\mathrm{o}}$ is the limit of the STWP spectrum [Fig.~\ref{Fig:SideBaseAngles}(a)]. For the sideband STWP, the spectral support is the intersection of the light-cone with the plane $\Omega\!=\!(k_{z}+k_{\mathrm{o}})\widetilde{v}$, and we take the non-differentiable wavelength to be $\lambda_{\mathrm{o}}\!=\!5000$~nm [Fig.~\ref{Fig:SideBaseAngles}(b)]. In this case, $\omega\!=\!\omega_{\mathrm{o}}$ corresponds to $k_{z}\!=\!-k_{\mathrm{o}}$. However, negative values of $k_{z}$ are excluded because they conflict with causal excitation and propagation \cite{Yessenov19PRA,Heyman87JOSAA}, and $\lambda_{\mathrm{o}}$ cannot belong to the STWP spectrum. Indeed, a whole angular span in the vicinity of $\varphi\!=\!0$ is excluded from sideband STWPs on physical grounds; specifically, the span of spatial frequencies $0\!<\!|k_{x}|\!<\!k_{\mathrm{o}}(1+\tfrac{\widetilde{v}}{c})$ is excluded.

We plot $\varphi(\omega)$ in Fig.~\ref{Fig:SideBaseAngles} for both of these STWPs, and two distinctions are immediately clear from Fig.~\ref{Fig:SideBaseAngles}. First, $\varphi$ is extremely large for the sideband STWP  with $\varphi_{\mathrm{c}}\!\approx\!49^{\circ}$ at $\lambda_{\mathrm{c}}\!=\!799$~nm, whereas $\varphi_{\mathrm{c}}\!\approx\!0.52^{\circ}$ for the baseband STWP ($\varphi_{\mathrm{o}}\!=\!0$ at $\lambda_{\mathrm{o}}\!=\!800$~nm). Therefore, the sideband STWP cannot be realized in the paraxial domain. Indeed, paraxial sideband STWPs necessitate using $\theta\!\rightarrow\!45^{\circ}$ ($\widetilde{v}\!\rightarrow\!c$), and $\omega_{\mathrm{o}}\!\rightarrow\!0$, in which case the structure is basically separable for narrow bandwidths. Second, the angular bandwidth is extremely narrow for the sideband STWP ($\Delta\varphi\!\approx\!0.06^{\circ}$) compared to that for its baseband counterpart ($\Delta\varphi\!\approx\!0.8^{\circ}$) for the same bandwidth $\Delta\lambda\!=\!2$~nm. Therefore, we expect sideband STWPs to be more sensitive to errors in the synthesis process. Both of these features are a consequence of the location of the selected spectrum with respect to the non-differentiable wavelength $\lambda_{\mathrm{o}}$.

We place these observation on a quantitative basis in Fig.~\ref{Fig:Error}. First, we calculate the lowest-order AD coefficients evaluated at $\lambda_{\mathrm{c}}$: $\varphi_{\mathrm{c}}$, $\varphi_{\mathrm{c}}^{(1)}\!=\!\tfrac{d\varphi}{d\omega}\big|_{\omega_{\mathrm{c}}}$, and $\varphi_{\mathrm{c}}^{(2)}\!=\!\tfrac{d^{2}\varphi}{d\omega^{2}}|_{\omega_{\mathrm{c}}}$ [Fig.~\ref{Fig:Error}(a)] for the sideband and the baseband STWPs from Fig.~\ref{Fig:SideBaseAngles}. We see that $\varphi_{\mathrm{c}}^{(1)}$ and $\varphi_{\mathrm{c}}^{(2)}$ for the baseband STWP are significantly larger ($\times80$ and $\times10^{3}$, respectively) than those for the sideband STWP. These AD coefficients determine the axial group velocity $\widetilde{v}$ and GVD coefficient $k_{2}\!=\!\tfrac{d^{2}k_{z}}{d\omega^{2}}\big|_{\omega_{\mathrm{c}}}$ of the two STWPs: $\widetilde{v}\!=\!\tfrac{c}{\cos{\varphi_{\mathrm{c}}}-\omega_{\mathrm{c}}\varphi_{\mathrm{c}}^{(1)}\sin{\varphi_{\mathrm{c}}}}$ and \cite{Porras03PRE2}:
\begin{equation}
c\omega_{\mathrm{c}}k_{2}=-(\omega_{\mathrm{c}}\varphi_{\mathrm{c}}^{(1)})^{2}\cos{\varphi_{\mathrm{c}}}-(\omega_{\mathrm{c}}^{2}\varphi_{\mathrm{c}}^{(2)}+2\omega_{\mathrm{c}}\varphi_{\mathrm{c}}^{(1)})\sin{\varphi_{\mathrm{c}}}.
\end{equation}
Although the values of the AD coefficients are quite different for the two STWPs [Fig.~\ref{Fig:Error}(a)], their axial group velocities and GVD coefficients are nevertheless identical [Fig.~\ref{Fig:Error}(b)], with $\widetilde{v}\!\approx\!1.19c$ and $k_{2}\!=\!0$. In other words, the two STWPs propagate invariantly at the same group velocity, as long as the AD coefficients are implemented exactly. 

Because $\varphi_{\mathrm{c}}^{(1)}$ and $\varphi_{\mathrm{c}}^{(2)}$ are significantly larger for the baseband STWP than for the sideband STWP, a high degree of precision is therefore needed to realize the sideband STWP. This can be demonstrated directly by adding a small error to the AD coefficients for both STWPs. Given the widespread use of spatial light modulators (SLMs) in synthesizing spatio-temporally structured fields, we consider the smallest error that is unavoidable in an SLM due to phase quantization. We estimate that this can produce a deviation of $\sim\!0.3\%$ in the produced angle $\varphi(\lambda)$. We plot the new AD coefficients in Fig.~\ref{Fig:Error}(c) for the two STWPs after incorporating this error. The shift in values is extremely small ($\omega_{\mathrm{c}}\delta\varphi_{\mathrm{c}}^{(1)}\!\sim\!0.011$ and 0.003, and $\omega_{\mathrm{c}}^2\delta\varphi_{\mathrm{c}}^{(2)}\!\sim\!4.456$ and $0.008$ for the baseband and sideband STWPs, respectively). The values of $\widetilde{v}$ and $k_{2}$ for the sideband STWP undergo a significant change, whereas those for the baseband STWP are almost unchanged. In particular, the sideband STWP experiences very high GVD in free space, in excess of that for a highly dispersive medium such as ZnSe. The sideband STWP is therefore no longer propagation invariant in free space after incorporating this minute error.

These results are quite generic and apply for other values of $\theta$ and central wavelength $\lambda_{\mathrm{c}}$. In all cases, baseband STWPs are more robust than sideband STWPs with respect to changes occurring in the realized group velocity and GVD coefficients upon deviation in the implemented AD coefficients from the ideal values. This conclusion emphasizes the difficulty in producing sideband STWPs, which is reinforced by the lack of experimental realizations of such pulsed beams over the past $\sim\!40$~years. Baseband STWPs whose spectra are in the vicinity of the non-differentiable wavelength, on the other hand, are free of such tight restrictions, which is exemplified by the rapid experimental progress over the past few years sine they have been discovered \cite{Yessenov22AOP}.

\section{The Schmidt number as an estimate of the accessible degrees of freedom}

\subsection{Impact of the spectral uncertainty}

An unavoidable impairment in realizing any AD profile (associated with an STWP or a TPF) is the `spectral uncertainty' \cite{Yessenov19OE,Kondakci19OL}: each angle $\varphi$ is associated with a finite bandwidth $\delta\omega$ centered on $\omega$, rather than being strictly associated with the single frequency $\omega$. In other words, the ideal conception of AD requires a delta-function correlation between the frequency $\omega$ and the angle $\varphi$. The spectral uncertainty is an inevitable result of finite experimental resources; e.g., the spectral resolution of finite-sized diffraction gratings, and the finite aperture and pixel size of SLMs \cite{Kondakci19OL,Yessenov19OE}. The analysis so far has been idealized by assuming $\delta\omega\!\rightarrow\!0$; as such, the spectral support on the light-cone is strictly a 1D trajectory [Fig.~\ref{Fig:LightCones}(c,d)], and the field has \textit{infinite} energy whether it is a TPF or an STWP \cite{Sezginer85JAP,Porras03PRE,Yessenov19OE}. In any real system, relaxing the delta-function association between $\omega$ and $\varphi$ renders the pulse energy finite. 

\begin{figure*}[t!]
\centering
\includegraphics[width=17.6cm]{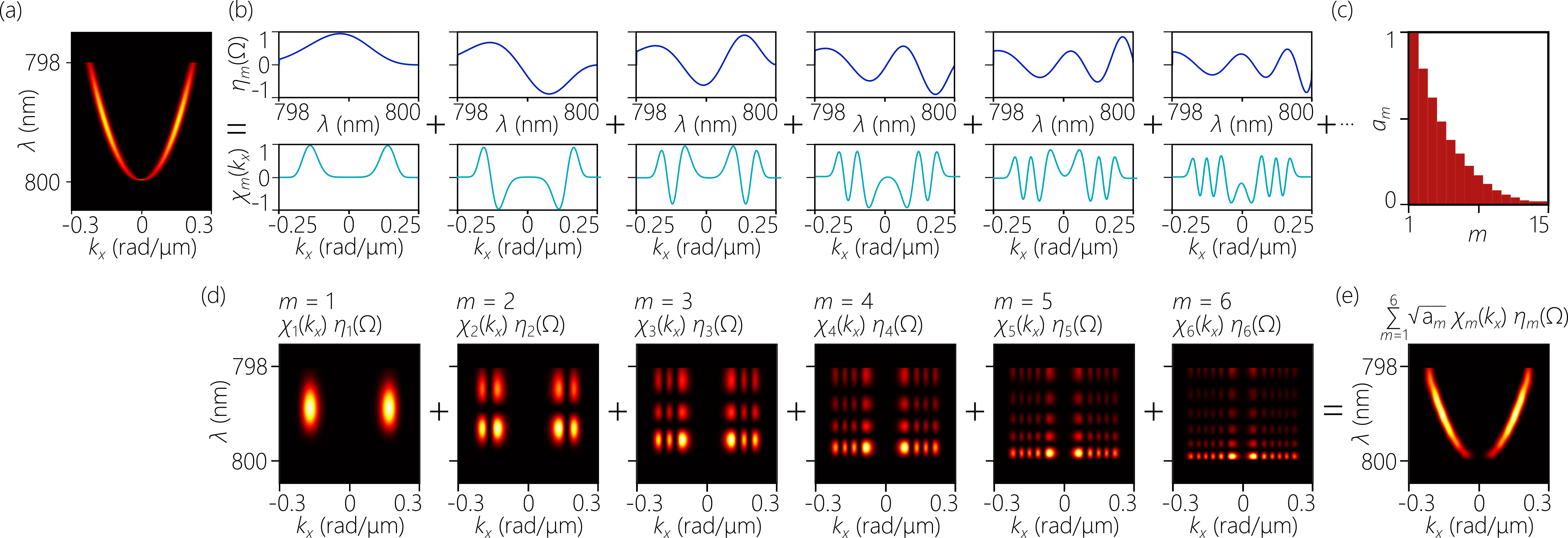}
\caption{(a) Spatio-temporal spectrum $|\widetilde{\psi}(k_{x},\lambda)|^{2}$ for a abseband STWP with $\theta\!=\!50^{\circ}$, $\lambda_{\mathrm{o}}\!=\!800$~nm, $\lambda_{\mathrm{c}}\!=\!799$~nm, $\delta\lambda\!=\!10$~pm, and $\Delta\lambda\!=\!2$~nm. (b) The first 6 Schmidt modes from the sets $\{\chi_{m}(k_{x})\}$ and $\{\eta_{m}(\lambda)\}$. (c) Schmidt weights $\{a_{m}^{2}\}$ normalized with respect to the first coefficient. (d) Composite Schmidt modes $|\chi_{m}(k_{x})\eta(\Omega)|^{2}$ for $m\!=\!1,\cdots,6$. (e) The spatio-temporal spectrum $|\widetilde{\psi}(k_{x},\lambda)|^{2}\!=\!|\sum_{m=1}^{6}a_{m}\chi_{m}(k_{x})\eta_{m}(\lambda)|^{2}$ with equal weights $a_{m}\!=\!\tfrac{1}{\sqrt{6}}$.}
\label{Fig:SchmidtModes}
\end{figure*}

\begin{figure}[t!]
\centering
\includegraphics[width=8.6cm]{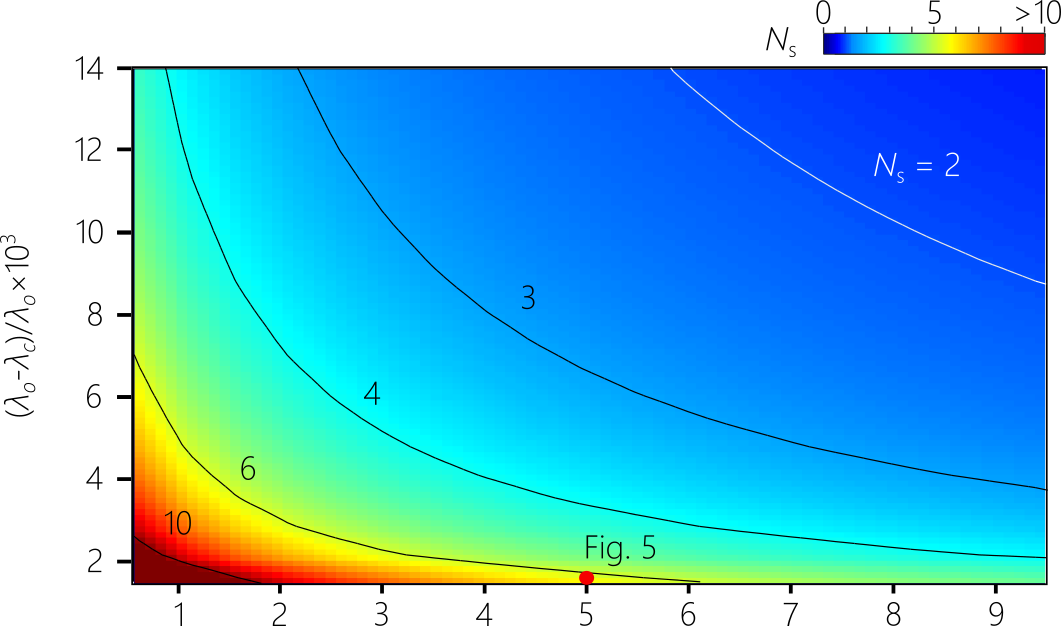}
\caption{Calculated Schmidt number $N_{\mathrm{S}}$ for the spatio-temporal spectrum of a STWP at $\theta\!=\!50^{\circ}$ and of bandwidth $\Delta\lambda\!=\!2$~nm as a function of the normalized spectral uncertainty and the normalized spectral separation between the non-differentiable wavelength $\lambda_{\mathrm{o}}$ and the central wavelength $\lambda_{\mathrm{c}}$. The Schmidt number increases as we approach the ideal condition (lower left corner): vanishing spectral uncertainty and approaching the non-differentiable frequency. The red dot identifies the condition for the STWP shown in Fig.~\ref{Fig:SchmidtModes}.}
\label{Fig:SchmidtNumberST}
\end{figure}

We proceed to quantify how the spectral uncertainty in the AD profile of an STWP curtails control over its characteristics (e.g., the group velocity and GVD coefficient), and how the proximity to the non-differentiable frequency $\omega_{\mathrm{o}}$ is beneficial in minimizing this deleterious effect. To include the impact of the spectral uncertainty into the STWP envelope, we modify Eq.~\ref{Eq:SWTPEnvelope} as follows:
\begin{equation}
\psi(x,z;t)=\iint\!d\Omega dk_{x}\widetilde{\psi}(\Omega)\widetilde{g}(k_{x}-k_{x,\mathrm{o}}(\Omega))e^{i\{k_{x}x+(k_{z}-k_{\mathrm{o}})z-\Omega t\}},
\end{equation}
where the integral is now over the temporal frequency $\Omega$ \textit{and} the spatial frequency $k_{x}$. The newly introduced function $\widetilde{g}(k_{x})$ is narrow, centered at $k_{x}\!=\!0$, and its width corresponds to the angular spread associated with each frequency $\omega$. The function $\widetilde{g}(k_{x})$ is shifted at each frequency to the value $k_{x,\mathrm{o}}(\Omega)\!=\!k_{\mathrm{o}}\sqrt{2(1-\cot{\theta})\tfrac{\Omega}{\omega_{\mathrm{o}}}}$ that would be realized in absence of spectral uncertainty. The energy of the wave packet in any axial plane, $\iint\!dxdt\;|\psi(x,z;t)|^{2}$, is now finite; however, the resulting STWP is no longer propagation invariant. Instead, the STWP is invariant for a distance $L_{\mathrm{max}}\!\sim\!\tfrac{c}{\delta\omega|1-\cot{\theta}|}$, after which it diffracts rapidly \cite{Kondakci16OE,Kondakci19OL,Yessenov19OE}.

As long as $\varphi$ has a one-to-one relationship with $\omega$ (i.e., $\delta\omega\!\rightarrow\!0$), there is in principle no limit on the control that may be exercised on the STWP characteristics. For example, considering propagation-invariant STWPs with $k_{2}\!=\!0$, it should be possible to distinguish any pair of STWPs no matter how close the values of their group velocities. However, a finite spectral uncertainty $\delta\omega$ results in an ambiguity in STWPs with close values of $\widetilde{v}$. Similarly, for two STWPs with the same $\widetilde{v}$ but different GVD coefficients, the finite spectral uncertainty sets a limit on the minimal difference in $k_{2}$ that can be delineated. To quantify the reduction in control over the STWP characteristics as a consequence of finite spectral uncertainty, we make use of the Schmidt decomposition as a quantitative measure \cite{Schmidt06MA}.

\subsection{Schmidt decomposition of the spatio-temporal spectrum}

There are two limits on the spatio-temporal spectrum $\widetilde{\psi}(k_{x},\Omega)\!=\!\widetilde{\psi}(\Omega)\widetilde{g}(k_{x}-k_{x}(\Omega))$: (1) $\widetilde{g}(k_{x})\!\rightarrow\!\delta(k_{x})$, whereupon we retrieve the perfect one-to-one relationship between $k_{x}$ and $\omega$ that is characteristic of ideal AD, in which case it is expected that arbitrary control can be exercised over the STWP characteristics; and (2) $\delta\omega\!\rightarrow\!\Delta\omega$, whereupon $\widetilde{\psi}(k_{x},\Omega)$ becomes separable with respect to the spatial and temporal frequencies, $\widetilde{\psi}(k_{x},\Omega)\!\rightarrow\!\widetilde{\psi}_{x}(k_{x})\widetilde{\psi}_{t}(\Omega)$, and the STWP tends to a conventional pulsed beam (Eq.~\ref{Eq:GeneralEnvelope}). We expect that the spatio-temporal spectrum in the first limit is associated with a large number of degrees of freedom, whereas the second limit reduces it to a single degree of freedom corresponding to the separable product of $\widetilde{\psi}_{x}(k_{x})$ and $\widetilde{\psi}_{t}(\Omega)$. These two limits suggest the suitability of the Schmidt decomposition for $\widetilde{\psi}(k_{x},\Omega)$ to determine its proximity to either of these two limits. Schmidt analysis \cite{Schmidt06MA} has been used extensively in quantum optics to analyze the entanglement in the spectrum of two-photon states \cite{Ekert95AJP,Law00PRL,Law04PRL,Eberly06LP,Borshchevskaia19LPL}. In such scenarios, the same physical degree of freedom (e.g., the temporal frequency $\omega$ \cite{Law00PRL} \textit{or} the spatial frequency $k_{x}$ \cite{Law04PRL}) is considered for each photon partaking in the two-photon state. In our case here, however, we have an altogether different situation where we consider two distinct physical degrees of freedom (the spatial and temporal frequencies).

For a finite spectral uncertainty, $\widetilde{\psi}(k_{x},\Omega)$ is square-integrable, and it thus admits a unique decomposition of the form:
\begin{equation}
\widetilde{\psi}(k_{x},\Omega)=\sum_{m}\sqrt{a_{m}}\chi_{m}(k_{x})\eta_{m}(\Omega),
\end{equation}
where $\{\chi_{m}(k_{x})\}$ and $\{\eta_{m}(\Omega)\}$ are two sets of orthonormal functions, a single index $m$ runs over both modal bases in the decomposition, and $\{a_{m}\}$ are real numbers normalized such that $\sum_{m}a_{m}\!=\!\iint dxdt|\psi(x,z;t)|^{2}\!=\!1$ for all $z$. This decomposition is the continuous-function equivalent of the well-known singular-value decomposition from matrix algebra. The set $\{\chi_{m}(k_{x})\}$ comprises the eigenfunctions of the Hermitian operator $\chi(k_{x},k_{x}')$ defined as:
\begin{equation}
\chi(k_{x},k_{x}')=\int\!\widetilde{\psi}(k_{x},\Omega)\widetilde{\psi}^{*}(k_{x}',\Omega)d\Omega=\sum_{m}a_{m}\chi_{m}(k_{x})\chi_{m}^{*}(k_{x}'),
\end{equation}
such that
\begin{equation}
\int\!\chi(k_{x},k_{x}')\chi_{m}(k_{x}')dk_{x}'=a_{m}\chi_{m}(k_{x}),
\end{equation}
where $\{a_{m}\}$ are the eigenvalues. Similarly, the set $\{\eta_{m}(\Omega)\}$ comprises the eigenfunctions of the Hermitian operator $\eta(\Omega,\Omega')$ defined as:
\begin{equation}
\eta(\Omega,\Omega')=\int\!\widetilde{\psi}(k_{x},\Omega)\widetilde{\psi}^{*}(k_{x},\Omega')dk_{x}=\sum_{m}a_{m}\chi_{m}(\Omega)\chi_{m}^{*}(\Omega'),
\end{equation}
such that
\begin{equation}
\int\!\eta(\Omega,\Omega')\eta_{m}(\Omega')d\Omega'=a_{m}\eta_{m}(\Omega)
\end{equation}
where $\{a_{m}\}$ are once again the eigenvalues.

It is common to introduce a Schmidt number $N_{\mathrm{S}}$:
\begin{equation}
N_{\mathrm{S}}=\frac{1}{\sum_{m}a_{m}^{2}},
\end{equation}
which gives an effective number of spatio-temporal mode pairs (one for $k_{x}$ and the other for $\Omega$) required to construct $\widetilde{\psi}(k_{x},\Omega)$. For example, if the modal coefficients are equal $a_{m}\!=\!\tfrac{1}{\sqrt{M}}$ for $m\!=\!1,2,\cdots,M$ and $a_{m}\!=\!0$ otherwise, then $N_{\mathrm{S}}\!=\!M$. If only one term exists in the Schmidt decomposition, then $N_{\mathrm{S}}\!=\!1$ and the spectrum is separable with respect to $k_{x}$ and $\Omega$, so that the field is AD-free. In presence of perfect association between $\Omega$ and $k_{x}$, which corresponds to ideal AD (i.e., $\delta\omega\!\rightarrow\!0$), $N_{\mathrm{S}}\!\rightarrow\!\infty$. 

We recently made use of Schmidt analysis to analyze the mutual intensity of the field $J(x_{1},x_{2};z)\!=\!int\!dt\,\psi(x_{1},z;t)\psi^{*}(x_{2},z;t)$ to obtain the dependence of the propagation distance $L_{\mathrm{max}}$ of STWPs on the spectral uncertainty \cite{Kondakci19OL}. In \cite{Kondakci19OL} we carried the Schmidt number to represent the degree of classical entanglement \cite{Qian11OL,Kagalwala13NP} between the continuous spatial and temporal degrees of freedom. More work is needed determine the relationship between the Schmidt number in \cite{Kondakci19OL} and that calculated here.

\begin{figure*}[t!]
\centering
\includegraphics[width=14cm]{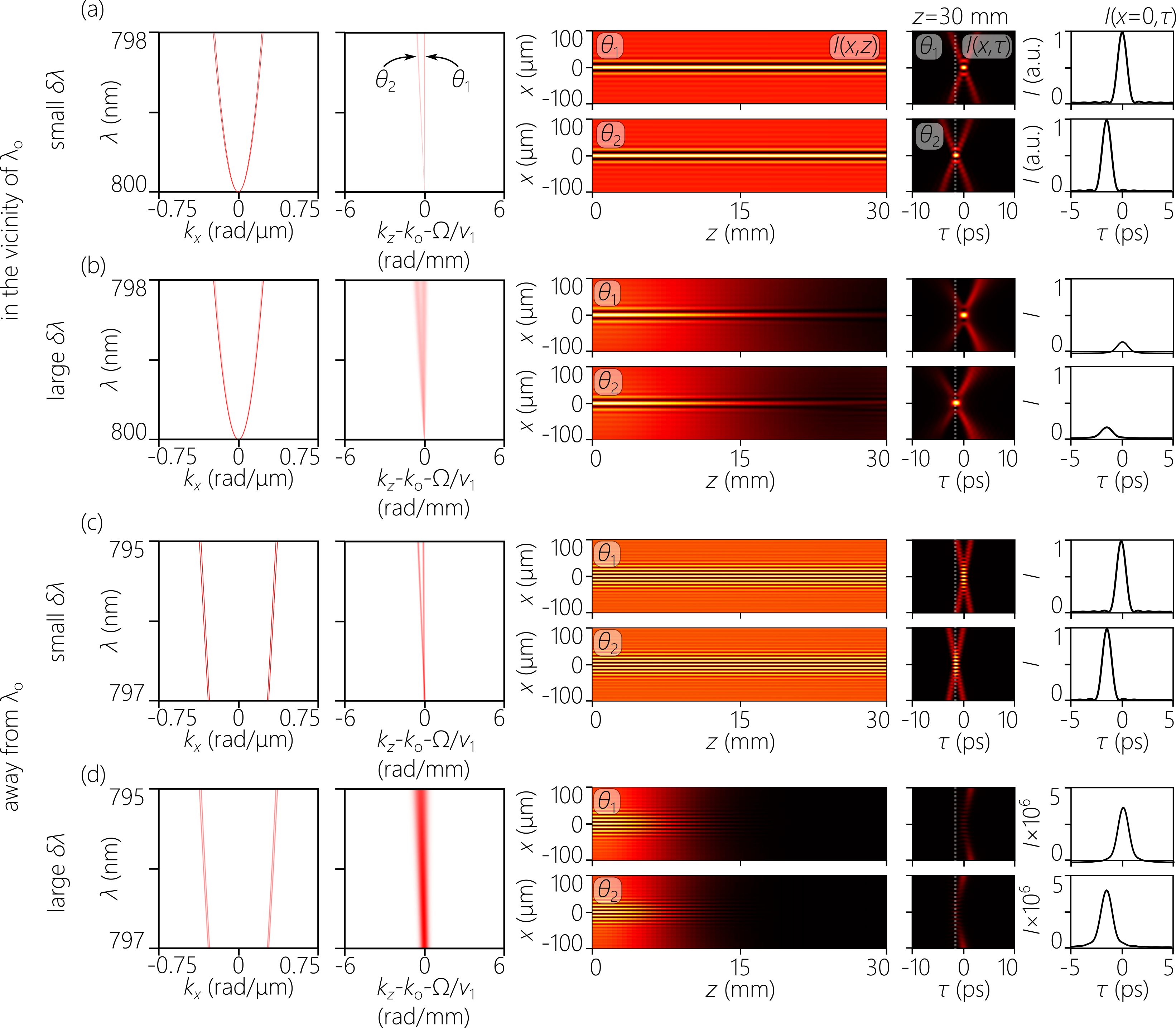}
\caption{Impact of the spectral uncertainty $\delta\lambda$ and proximity from the non-differentiable wavelength $\lambda_{\mathrm{o}}$ on the control over the STWP group velocity $\widetilde{v}$. In each row, the plots from left to right are: the spectral projection onto the $(k_{x},\lambda)$-plane; the second-column onto the $(k_{z},\lambda)$-plane; the time-averaged intensity $I(x,z)$; the spatio-temporal intensity profile $I(x,\tau)$ at $z\!=\!30$~mm, where $\tau\!=t-z/(c\tan{\theta_{1}}$; and the on-axis ($x=0$) temporal profile $I(\tau)$. The calculations are carried out for two propagation-invariant STWPs with $\theta_{1}\!=\!50^\circ$ and $\theta_{2}\!=\!50.5^{\circ}$, and both have $\Delta\lambda\!=\!2$~nm and $\lambda_{\mathrm{o}}\!=\!800$~nm. (a,b) The STWP spectra include $\lambda_{\mathrm{o}}$ ($\lambda_{\mathrm{c}}\!=\!799$~nm); in (a) $\delta\lambda\!=\!10$~pm and in (b) $\delta\lambda\!=\!100$~pm. (c,d) Same as (a,b), except that $\lambda_{\mathrm{c}}\!=\!796$~nm (the spectra do not include $\lambda_{\mathrm{o}}$).}
\label{Fig:PropagationDistance}
\end{figure*}

\begin{figure*}[t!]
\centering
\includegraphics[width=14cm]{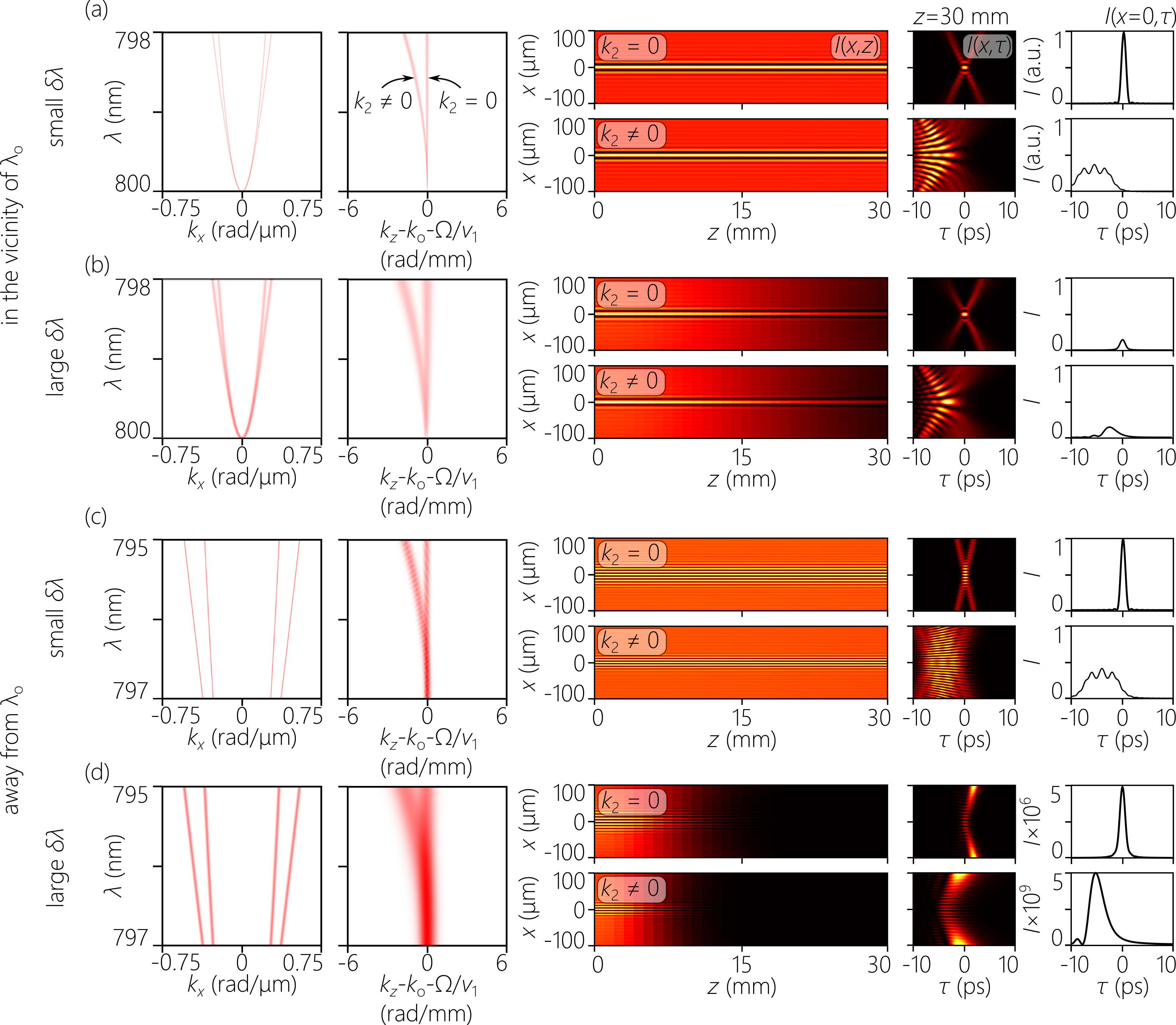}
\caption{Impact of the spectral uncertainty $\delta\lambda$ and proximity from the non-differentiable wavelength $\lambda_{\mathrm{o}}$ on the control over the GVD coefficient $k_{2}$ of an STWP in free space. In each row, the plots from left to right are the same as in Fig.~\ref{Fig:PropagationDistance}. The calculations are carried out for two STWPs with $\theta\!=\!50^\circ$, $\Delta\lambda\!=\!2$~nm, and $\lambda_{\mathrm{o}}\!=\!800$~nm. One STWP is propagation invariant with $k_{2}\!=\!0$ and the other STWP is dispersive with $\omega_{\mathrm{o}}ck_{2}\!=\!-100$. (a,b) The STWP spectra include $\lambda_{\mathrm{o}}$ ($\lambda_{\mathrm{c}}\!=\!799$~nm); in (a) $\delta\lambda\!=\!10$~pm and in (b) $\delta\lambda\!=\!100$~pm. (c,d) Same as (a,b), except that $\lambda_{\mathrm{c}}\!=\!796$~nm (the spectra do not include $\lambda_{\mathrm{o}}$).}
\label{Fig:SchmidtNumberDispersion}
\end{figure*}

\subsection{Schmidt decomposition for an STWP}

We consider a concrete example in Fig.~\ref{Fig:SchmidtModes} of an STWP having bandwidth $\Delta\lambda\!=\!2$~nm, spectral uncertainty $\delta\lambda\!=\!10$~pm, non-differentiable wavelength $\lambda_{\mathrm{o}}\!=\!800$~nm, and a spectral tilt angle $\theta\!=\!50^{\circ}$; the spatio-temporal spectrum is plotted in Fig.~\ref{Fig:SchmidtModes}(a). We compute the bi-orthogonal bases $\{\chi_{m}(k_{x})\}$ and $\{\eta_{m}(\Omega)\}$ for the Schmidt decomposition associated with this STWP and plot the first 6 modes in each set in Fig.~\ref{Fig:SchmidtModes}(b), and plot the Schmidt coefficients $\{a_{m}\}$ in Fig.~\ref{Fig:SchmidtModes}(c). We note a crucial difference between the bi-orthogonal basis sets for the spatial and temporal frequencies: whereas the modes in $\{\eta_{m}(\Omega)\}$ resemble Hermite-Gaussian modes, the mirror-symmetry around $k_{x}\!=\!0$ results in the modes $\{\chi_{m}(k_{x})\}$ resembling pairs of Hermite-Gaussian modes. The Schmidt number for this STWP in $N_{\mathrm{S}}\!\approx\!7.2$. By truncating the Schmidt decomposition over the first 6 terms [Fig.~\ref{Fig:SchmidtModes}(d)], the resulting form of the spatio-temporal spectrum does indeed match the original spectrum except in the vicinity of the non-differentiable wavelength $\lambda_{\mathrm{o}}\!=\!800$~nm where a dip forms, which `fills up' only with the addition of higher-order terms in the Schmidt sum. This already indicates that the main contribution of the higher-order terms is in the vicinity of $\lambda_{\mathrm{o}}$.

This conclusion is confirmed in Fig.~\ref{Fig:SchmidtNumberST} where we plot the Schmidt number $N_{\mathrm{S}}$ for a baseband STWP ($\theta\!=\!50^{\circ}$ and $\Delta\lambda\!=\!2$~nm) while varying two key parameters: the spectral uncertainty $\delta\omega$ and the displacement of the central frequency $\omega_{\mathrm{c}}$ from the non-differentiable frequency $\omega_{\mathrm{o}}$. The results in Fig.~\ref{Fig:SchmidtNumberST} help establish several trends. As expected from previous literature \cite{Law00PRL}, the spectral uncertainty decreases the number of degrees of freedom drastically for any $\omega_{\mathrm{c}}$. Furthermore, as $\omega_{\mathrm{c}}$ moves away from $\omega_{\mathrm{o}}$, $N_{\mathrm{S}}$ decays exponentially down to a baseline of $N_{\mathrm{S}}\!\rightarrow\!1$. The impact of these two factors (increasing the spectral uncertainty and the separation from $\omega_{\mathrm{o}}$) is to reduce $N_{\mathrm{S}}$, which in turn reduces our control over the STWP propagation characteristics as we proceed to demonstrate.

\subsection{Impact of the spectral uncertainty on the tunability of the group velocity}

We first consider the impact of the spectral uncertainty alongside the proximity to the non-differentiable frequency $\omega_{\mathrm{o}}$ on the tunability of the group velocity of two propagation-invariant baseband STWPs ($k_{2}\!=\!0$). We take the spectral tilt angles of the two STWPs to be $\theta_{1}\!=\!50^{\circ}$ and $\theta_{2}\!=\!50.5^{\circ}$, corresponding to a difference in group velocity of $\Delta\widetilde{v}\!\approx\!0.02c$. Both STWPs have a bandwidth of $\Delta\lambda\!=\!2$~nm and a non-differentiable wavelength of $\lambda_{\mathrm{o}}\!=\!800$~nm.

We consider first the impact of the spectral uncertainty alone and hold the central frequency of the spectrum fixed at $\lambda_{\mathrm{c}}\!=\!799$~nm, so that the spectra for both STWPs include the non-differentiable wavelength $\lambda_{\mathrm{o}}$. We change the spectral uncertainty from $\delta\lambda\!=\!10$~pm (small $\delta\lambda$) whereupon $N_{\mathrm{S}}\!\approx\!7.2$, to $\delta\lambda\!=\!100$~pm (large $\delta\lambda$), whereupon $N_{\mathrm{S}}\!\approx\!3.4$, and plot the results in Fig.~\ref{Fig:PropagationDistance}(a) and Fig.~\ref{Fig:PropagationDistance}(b), respectively. The impact of $\delta\lambda$ on the distinguishability of the spectra for the two STWPs is particularly clear in the spectral projection onto the $(k_{z},\lambda)$-plane. The spectra are distinguishable in the case of low $\delta\lambda$, and the on-axis ($x\!=\!0$) pulses are separated by $\approx\!1.5$~ps after an axial propagation distance of 30~mm as expected. Note that $L_{\mathrm{max}}\!>\!30$~mm when $\delta\lambda\!=\!10$~pm, so that the pulses are recorded at $z\!=\!30$~mm with no drop in intensity. The spectra start to overlap in the case of large $\delta\lambda$. However, the propagation distance is reduced $L_{\mathrm{max}}\!<\!30$~mm, so that the on-axis pulse intensities in Fig.~\ref{Fig:PropagationDistance}(b) have dropped with respect to those in Fig.~\ref{Fig:PropagationDistance}(a), in addition to some pulse distortion, after a propagation distance of 30~mm.

We next consider two similar baseband STWPs except that $\lambda_{\mathrm{c}}\!=\!796$~nm; i.e., the spectra have been shifted by 3~nm away from the non-differentiable wavelength $\lambda_{\mathrm{o}}\!=\!800$~nm [Fig.~\ref{Fig:PropagationDistance}(c,d)]. The Schmidt numbers for these two STWPs are $N_{\mathrm{S}}\!\approx\!3.2$ (small $\delta\lambda$) and $N_{\mathrm{S}}\!\approx\!1.5$ (large $\delta\lambda$). Because the spatial spectra are narrower and shifted from $k_{x}\!=\!0$, the spatial profiles here are broader and feature a sinusoidal modulation that is absent from the two STWPs in Fig.~\ref{Fig:PropagationDistance}(a,b). The shift away from the non-differentiable wavelength $\lambda_{\mathrm{o}}$ has a clear deleterious impact: the same spectral uncertainty $\delta\lambda$ broadens the spectral projection onto the $(k_{z},\lambda)$-plane. The relatively large $N_{\mathrm{S}}$ for small $\delta\lambda$ indicates that the target group velocities are still expected to be realized [Fig.~\ref{Fig:PropagationDistance}(c)]. However, the extremely small $N_{\mathrm{S}}$ associated with large $\delta\lambda$ is associated with considerable overlap between the spectra for the two STWPs and on-axis pulse deformation, and the reduced $L_{\mathrm{max}}$ results in an almost extinction of the STWPs at $z\!=\!30$~mm [Fig.~\ref{Fig:PropagationDistance}(d)].

\subsection{Impact of the spectral uncertainty on the tunability of the GVD coefficient}

A similar impact of the spectral uncertainty and proximity from the non-differentiable wavelength $\lambda_{\mathrm{o}}$ is clear in Fig.~\ref{Fig:SchmidtNumberDispersion} with regards to control over the GVD coefficient $k_{2}$. We consider here two baseband STWPs with the same spectral tilt angle $\theta\!=\!50^{\circ}$. However, one STWP is propagation invariant with $k_{2}\!=\!0$, whereas the other is endowed with an anomalous GVD coefficient $c\omega_{\mathrm{o}}k_{2}\!=\!-100$. When the STWP spectrum includes $\lambda_{\mathrm{o}}$ [Fig.~\ref{Fig:SchmidtNumberDispersion}(a,b)] and the Schmidt number is large ($N_{\mathrm{S}}\!\approx\!9.2$ for small $\delta\lambda$ and $N_{\mathrm{S}}\!\approx\!4.5$ for large $\delta\lambda$), the two spectra for the propagation-invariant and the dispersive STWPs are readily distinguished. Whereas the expected dispersive pulse broadening is observed for small $\delta\lambda$ [Fig.~\ref{Fig:SchmidtNumberDispersion}(a)], this is not observed for large $\delta\lambda$ [Fig.~\ref{Fig:SchmidtNumberDispersion}(b)]. Nevertheless, the reduced $L_{\mathrm{max}}$ for large $\delta\lambda$ diminishes the pulse intensities at $z\!=\!30$~mm. Displacing the spectrum away from $\lambda_{\mathrm{o}}$ reduces the Schmidt numbers ($N_{\mathrm{S}}\!\approx\!6.5$ for small $\delta\lambda$ and $N_{\mathrm{S}}\!\approx\!3.2$ for large $\delta\lambda$). Whereas the target GVD is approximately realized for small $\delta\lambda$ [Fig.~\ref{Fig:SchmidtNumberDispersion}(c)], this is not the case for large $\delta\lambda$ [Fig.~\ref{Fig:SchmidtNumberDispersion}(d)]. Furthermore, the reduced $L_{\mathrm{max}}$ for large $\delta\lambda$ practically extinguishes the on-axis STWP at $z\!=\!30$~mm [Fig.~\ref{Fig:SchmidtNumberDispersion}(d)].

\section{Conclusion}

In conclusion, we have shown that non-differentiable AD can be viewed as a resource that allows us to exercise precise control over the attributes of pulsed optical fields endowed with this feature, such as STWPs. New propagation behaviors are thus unlocked if the AD profile includes a wavelength at which the derivative of the propagation angle is not defined; e.g., propagation invariance, tunability of the group velocity, and the possibility of engendering normal or anomalous GVD in free space -- all realized for on-axis pulses. Although these consequences of non-differentiable AD can be observed in principle regardless of the location of the selected STWP spectrum with respect to the non-differentiable wavelength $\lambda_{\mathrm{o}}$, we have nevertheless found that operating in the vicinity of $\lambda_{\mathrm{o}}$ has two critical advantages: (1) robustness with respect to synthesis errors; and (2) an enhanced level of control over the field characteristics in presence of spectral uncertainty. Non-differentiable AD can thus be viewed as a `resource' that is optimally harnessed by operating in the vicinity of $\lambda_{\mathrm{o}}$, a fact that favors relying on `baseband' STWPs rather than their `sideband' counterparts (such as FWMs and X-waves). Finally, we have demonstrated the utility of Schmidt analysis for elucidating the nature of non-differentiable AD, and have suggested the Schmidt number associated with the spatio-temporal spectrum of the STWP as a quantitative measure for assessing the non-differentiable AD resource.

\section*{ACKNOWLEDGMENTS}
U.S. Office of Naval Research (ONR) N00014-17-1-2458 and N00014-20-1-2789.

\section*{Disclosures}
The authors declare no conflicts of interest.

\section*{Data availability}
Data underlying the results presented in this paper are not publicly available at this time but may be obtained from the authors upon reasonable request.


\bibliography{diffraction}

\end{document}